\documentclass[conference,a4paper,10pt]{IEEEtran}

\usepackage{amsmath,amssymb}
\usepackage{graphicx,psfrag}
\usepackage{color}
\usepackage{cite}
\usepackage{subfigure}
\usepackage{url}
\usepackage{array}
\usepackage{enumerate}
\usepackage{amsthm}
\usepackage{comment}
\usepackage{float}
\usepackage{balance}
\usepackage{algorithm}
\usepackage{algorithmic}
\usepackage{balance}
\usepackage{subfigure}

\IEEEoverridecommandlockouts

\newtheorem{theorem}{Theorem}
\newtheorem{lemma}{Lemma}
\newtheorem{proposition}{Proposition}

\newtheorem{remark}{Remark}

\newtheorem{definition}{Definition}

\DeclareMathOperator*{\argmin}{arg\,min}

\begin{document}

\title{Privacy-Utility Management of Hypothesis Tests}

\author{\IEEEauthorblockN{Zuxing Li\IEEEauthorrefmark{1} and Tobias J. Oechtering\IEEEauthorrefmark{2}}
\IEEEauthorblockA{\IEEEauthorrefmark{1}CentraleSup\'{e}lec \& L2S, Paris, France}
\IEEEauthorblockA{\IEEEauthorrefmark{2}EECS, KTH Royal Institute of Technology, Stockholm, Sweden}
\thanks{The work has been supported by the Swedish Research Council (VR) within the CHIST-ERA project COPES under Grant 2015-06815 and project CLONE under Grant E0628201.}
}

\maketitle

\begin{abstract}
The trade-off of hypothesis tests on the correlated privacy hypothesis and utility hypothesis is studied. The error exponent of the Bayesian composite hypothesis test on the privacy or utility hypothesis can be characterized by the corresponding minimal Chernoff information rate. An optimal management protects the privacy by minimizing the error exponent of the privacy hypothesis test and meanwhile guarantees the utility hypothesis testing performance by satisfying a lower bound on the corresponding minimal Chernoff information rate. The asymptotic minimum error exponent of the privacy hypothesis test is shown to be characterized by the infimum of corresponding minimal Chernoff information rates subject to the utility guarantees.
\end{abstract}

\section{Introduction}
\label{section1}
Privacy-utility trade-offs have been studied in different contexts \cite{wyner1975,varodayan2011,mhanna2015,nadendla2016,yao2017,giaconi2018,liao2018,lei2014,ghosh2012}. The privacy leakage can be modeled as a statistical inference and measured by the mutual information \cite{wyner1975,varodayan2011,mhanna2015,giaconi2018,liao2018}, divergence \cite{nadendla2016}, differential privacy \cite{ghosh2012}, or variance \cite{lei2014}. Depending on the application, the utility measure can be the expectation of cost \cite{yao2017,giaconi2018,lei2014}, divergence \cite{nadendla2016,liao2018}, or data rate \cite{wyner1975,mhanna2015}. With the privacy and utility measures, the trade-off problem can be formulated as a worst case analysis \cite{varodayan2011,nadendla2016,giaconi2018,liao2018,lei2014} or zero-sum game \cite{yao2017}.

The asymptotic error exponent of the simple binary hypothesis test with i.i.d. observations in \cite{nadendla2016} is characterized by a Kullback-Leibler divergence under the Neyman-Pearson criterion \cite{cover2006} or a Chernoff information under the Bayesian criterion \cite{chernoff1952}. It was shown in \cite{leang1997} that the asymptotic error exponent of a Bayesian $m$-ary hypothesis test with i.i.d. observations is characterized by the minimal Chernoff information among all conditional probability distribution pairs. In more general cases, the observations depend on correlated hypotheses. The asymptotically optimal composite hypothesis tests on one of the correlated hypotheses were studied under the Neyman-Pearson criterion, e.g., the Hoeffding test based on a Kullback-Leibler divergence statistic \cite{hoeffding1965} and the mismatched test based on a more relaxed mismatched divergence statistic \cite{unnikrishnan2011}.

In this paper, we consider a novel privacy-utility trade-off scenario where the Bayesian composite hypothesis tests on the correlated privacy hypothesis and utility hypothesis are made based on the same sequence of random observations\footnote{A such scenario in practice is the smart meter privacy problem, where the smart meter readings consist of the utility information, e.g., the future energy consumption from the grid, as well as the consumers' privacy, e.g., their life styles, and can be used by the authorized data recipient, e.g., the energy provider or grid operator, to make a utility hypothesis test and an illegitimate privacy hypothesis test.}. Firstly we prove and show that the error exponents of the Bayesian composite hypothesis tests on the privacy hypothesis and utility hypothesis can be characterized by their corresponding minimal Chernoff information rates. We then study the optimal management which degrades the privacy hypothesis test while guarantees a certain utility hypothesis testing performance, and further show that the asymptotic minimum error exponent of the privacy hypothesis test can be characterized by the infimum of corresponding minimal Chernoff information rates. In this initial study, we develop the results by considering a binary utility hypothesis and a binary privacy hypothesis. The extension to $m$-ary hypothesis tests, $m>2$, is our future work. In the context of distributed hypothesis test, a similar problem has been studied in our previous work \cite{zuxing20142}, where the Bayesian risks are used to measure the hypothesis testing performances and an optimal privacy-constrained distributed hypothesis testing network design is characterized. In the context of smart meter privacy, we characterized the optimal privacy-preserving energy management with an adversarial hypothesis test \cite{zuxing2017}.

\begin{figure}
\centering
\includegraphics[scale=0.5]{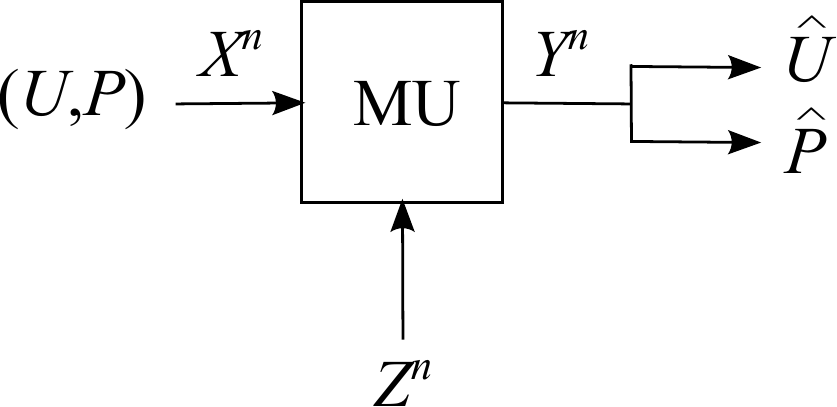}
\caption{Utility and privacy hypothesis tests based on the random observation sequence $Y^{n}$ processed by the management unit (MU), which has an i.i.d. random input sequence $X^{n}$ conditioned on a hypothesis pair realization $(u,p)$ and an i.i.d. random noise sequence $Z^{n}$.}
\label{figure1}
\end{figure}

\section{Problem Statement}
\label{section2}
The model in Fig. \ref{figure1} shows a management of hypothesis tests on the correlated privacy hypothesis and utility hypothesis in the presence of an independent noise. Let $U$ denote the binary utility hypothesis and $P$ denote the binary privacy hypothesis that has to be protected. W.l.o.g., we assume the following hypothesis alphabets $\mathcal{U}=\mathcal{P}=\{0,1\}$. Let $p_{U,P}$ denote the joint prior probability distribution of the random hypothesis pair. Given a hypothesis pair realization $(u,p)\in\mathcal{U}\times\mathcal{P}$, $X_{i}$ is i.i.d. generated following the pmf $p_{X|U=u,P=p}$, which will be denoted by $p_{X|u,p}$ in the following. We further assume that the four different pmfs $\{p_{X|u,p}\}_{(u,p)\in\mathcal{U}\times\mathcal{P}}$ are defined on the same finite support set $\mathcal{X}$. An independent random noise $Z_{i}$ defined on the finite alphabet $\mathcal{Z}$ is i.i.d. generated following the pmf $p_{Z}$. Over an $n$-slot time horizon, the management unit (MU) employs a randomized management policy $\phi^{n}_s:\mathcal{X}^{n}\times\mathcal{Z}^{n}\to\mathcal{X}^{n}$, which maps the input sequence $x^{n}$ and noise sequence $z^{n}$ to a processed random observation sequence $Y^{n}\in\mathcal{X}^{n}$ subject to the following constraint\footnote{In the context of smart meter privacy problem, this constraint means that the energy consumption $x^{n}$ can always be satisfied by the main energy supply $y^{n}$ and the alternative energy supply $z^{n}$ while the rate of wasted energy is bounded.}:
\begin{equation}
0\leq\frac{1}{n}\sum_{i=1}^{n}y_{i}+z_{i}-x_{i}\leq s.
\label{eb1}
\end{equation}
Bayesian hypothesis tests on the utility hypothesis $U$ and the privacy hypothesis $P$ are made with the objective to minimize their error probabilities based on the processed random sequence $Y^{n}$.

There are two objectives in the management policy design: Enhance the utility hypothesis test; and degrade the privacy hypothesis test. The two objectives are generally conflicting to each other. Therefore, the trade-off of hypothesis tests needs to be studied in the management policy design.

\section{Error Exponent of Bayesian Composite Hypothesis Test}
\label{section3}
In this section, we characterize fundamental bounds on the error exponent of Bayesian composite hypothesis test. The results derived in this section serve as the basis for the remaining discussions.

\subsection{Asymptotic Error Exponent with I.I.D. Observations}
Let $\alpha_{U}(X^{n})$ and $\alpha_{P}(X^{n})$ denote the minimal error probabilities of the Bayesian composite hypothesis tests on $U$ and $P$ based on the i.i.d. sequence $X^{n}$, e.g., a deterministic management policy $\phi_{s}^{n}(x^{n},z^{n})=x^{n}$ is employed when $z^{n}$ is a deterministic sequence of zeros. The corresponding asymptotic error exponents are characterized in the following Theorem \ref{tc1}, which is an extension of Chernoff theorem \cite{chernoff1952} to the Bayesian composite hypothesis test. To this end, we first introduce a function $T(Q_{1}||Q_{2};Q_{3})$, which is the minimum Kullback-Leibler divergence $T(Q_{1}||Q_{2};Q_{3})=\min_{Q\in\mathcal{Q}}D(Q||Q_{2})$ with $\mathcal{Q}=\{Q:D(Q||Q_{1})\leq D(Q||Q_{2}),D(Q||Q_{1})\leq D(Q||Q_{3})\}$.

\begin{definition}
Given three pmfs $Q_{1}$, $Q_{2}$, and $Q_{3}$ with the same support set $\mathcal{S}$, we define
\begin{equation*}
\begin{gathered}
T_{\mu,\nu}(Q_{1}||Q_{2};Q_{3})=-\log\sum_{a\in\mathcal{S}}Q_{1}^{\mu+\nu}(a)Q_{2}^{1-\mu}(a)Q_{3}^{-\nu}(a),\\
T(Q_{1}||Q_{2};Q_{3})=\max_{1\geq\mu\geq0,\nu\geq0}T_{\mu,\nu}(Q_{1}||Q_{2};Q_{3}).
\end{gathered}
\end{equation*}
\label{dc1}
\end{definition}

Conceptually, the function $T$ is the extension for the Bayesian composite hypothesis test of the Chernoff information used in the standard Bayesian binary hypothesis test.

\begin{proposition}
\begin{equation}
\begin{aligned}
&\lim_{n\to\infty}\frac{1}{n}\log\frac{1}{\alpha_{U}(X^{n})}\\
&\;\;\;=\min_{u,p,\bar{p}\in\{0,1\}}\left\{T(p_{X|u,p}||p_{X|1-u,\bar{p}};p_{X|1-u,1-\bar{p}})\right\}.
\end{aligned}
\label{ec6}
\end{equation}
\label{lc1}
\end{proposition}

The proof of Proposition \ref{lc1} follows from Sanov's theorem \cite[Theorem 11.4.1]{cover2006} and is presented in the appendix. Likewise the result can be derived for the asymptotic error exponent of privacy hypothesis test with the i.i.d. sequence $X^{n}$. Let $C(Q_{1}||Q_{2})$ denote the Chernoff information of pmfs $Q_{1}$ and $Q_{2}$ as defined in \cite{chernoff1952} as
\begin{equation*}
C(Q_{1}||Q_{2})=\max_{1\geq\mu\geq0}-\log\sum_{a\in\mathcal{S}}Q_{1}^{\mu}(a)Q_{2}^{1-\mu}(a).
\end{equation*}

\begin{lemma}
Given pmfs $Q_{1}$, $Q_{2}$, and $Q_{3}$ with the same support set, we have
\begin{equation}
\min\left\{\begin{gathered}
T(Q_{1}||Q_{2};Q_{3})\\
T(Q_{1}||Q_{3};Q_{2})
\end{gathered}\right\}
=
\min\left\{\begin{gathered}
C(Q_{1}||Q_{2})\\
C(Q_{1}||Q_{3})
\end{gathered}\right\}.
\label{ec9}
\end{equation}
\label{lc2}
\end{lemma}

The proof of Lemma \ref{lc2} follows from the joint concavity of $T_{\mu,\nu}$ over $(\mu,\nu)\in\mathbb{R}^{2}$ and is presented in the appendix. The following theorem is a direct consequence of Proposition \ref{lc1} and Lemma \ref{lc2}, and shows that the asymptotic error exponents of Bayesian composite hypothesis tests on $U$ and $P$ with the i.i.d. sequence $X^{n}$ are characterized by their corresponding minimal Chernoff informations.

\begin{theorem}
\begin{equation}
\begin{aligned}
\lim_{n\to\infty}\frac{1}{n}\log\frac{1}{\alpha_{U}(X^{n})}
=&\,\min_{\bar{p},\tilde{p}\in\{0,1\}}\left\{C(p_{X|1,\bar{p}}||p_{X|0,\tilde{p}})\right\},\\
\lim_{n\to\infty}\frac{1}{n}\log\frac{1}{\alpha_{P}(X^{n})}
=&\,\min_{\bar{u},\tilde{u}\in\{0,1\}}\left\{C(p_{X|\bar{u},1}||p_{X|\tilde{u},0})\right\}.
\end{aligned}
\label{ec10}
\end{equation}
\label{tc1}
\end{theorem}

\begin{remark}
Theorem \ref{tc1} cannot be implied from the asymptotic error exponent of the Bayesian multiple hypothesis test \cite{leang1997} since the minimal Chernoff information corresponding to a Bayesian composite hypothesis test in (\ref{ec10}) is among four combinations of pmfs rather than all six combinations of pmfs. However, the proof ideas in \cite{leang1997} can be used to show the asymptotic error exponents of the Bayesian composite hypothesis tests in Theorem \ref{tc1}: Bound the minimal error probability of a composite hypothesis test by the minimal error probabilities of binary hypothesis tests; and then use the Chernoff theorem \cite{chernoff1952} to bound the asymptotic error exponent of the Bayesian composite hypothesis test by the Chernoff informations of the binary hypothesis tests.
\end{remark}

\subsection{Lower Bound on the Error Exponent}
The problem shown in Fig. \ref{figure1} considers hypothesis tests on the utility hypothesis and the privacy hypothesis based on the processed (not necessarily i.i.d.) sequence $Y^{n}$. When a randomized management policy $\phi^{n}_{s}$ is used, let $\alpha_{U}(Y^{n},\phi^{n}_{s})$ and $\alpha_{P}(Y^{n},\phi^{n}_{s})$ denote the minimal error probabilities of the Bayesian composite hypothesis tests on $U$ and $P$ based on $Y^{n}=\phi^{n}_{s}(X^{n},Z^{n})$. The following proposition gives lower bounds on the error exponents in terms of the corresponding Chernoff information rates when a management policy $\phi_{s}^{n}$ is used.

\begin{proposition}
Given a management policy $\phi^{n}_{s}$ and the resulting pmfs $p_{Y^{n}|0,0}$, $p_{Y^{n}|0,1}$, $p_{Y^{n}|1,0}$, $p_{Y^{n}|1,1}$, then we have
\begin{equation}
\begin{aligned}
\frac{1}{n}&\log\frac{1}{\alpha_{U}(Y^{n},\phi^{n}_{s})}\\
&\geq\frac{1}{n}\min_{\bar{p},\tilde{p}\in\{0,1\}}\left\{C(p_{Y^{n}|1,\bar{p}}||p_{Y^{n}|0,\tilde{p}})\right\}-\frac{\log8p_{\max}}{n},\\
\frac{1}{n}&\log\frac{1}{\alpha_{P}(Y^{n},\phi^{n}_{s})}\\
&\geq\frac{1}{n}\min_{\bar{u},\tilde{u}\in\{0,1\}}\left\{C(p_{Y^{n}|\bar{u},1}||p_{Y^{n}|\tilde{u},0})\right\}-\frac{\log8p_{\max}}{n},
\end{aligned}
\label{ed6}
\end{equation}
where $p_{\max}=\max_{u,p\in\{0,1\}}\{p_{U,P}(u,p)\}$.
\label{ld1}
\end{proposition}

Note that $p_{\max}\geq\frac{1}{4}$ and $\frac{\log8p_{\max}}{n}\geq\frac{\log2}{n}\geq0$. The proof of Proposition \ref{ld1} is given in the appendix.

\section{Optimal Management of Hypothesis Tests}
\label{section4}
We denote a management policy $\phi^{n}_{s}$ by $\phi^{n}_{s,\lambda}$ if the resulting pmfs $p_{Y^{n}|0,0}$, $p_{Y^{n}|0,1}$, $p_{Y^{n}|1,0}$, $p_{Y^{n}|1,1}$ jointly satisfy the following utility hypothesis testing performance guarantee:
\begin{equation}
\frac{1}{n}\min_{\bar{p},\tilde{p}\in\{0,1\}}\left\{C(p_{Y^{n}|1,\bar{p}}||p_{Y^{n}|0,\tilde{p}})\right\}\geq\lambda+\frac{\log8p_{\max}}{n}.
\label{ed4}
\end{equation}
From Proposition \ref{ld1}, it follows that a policy $\phi^{n}_{s,\lambda}$ also satisfies a guarantee on the error exponent of the utility hypothesis test:
\begin{equation}
\frac{1}{n}\log\frac{1}{\alpha_{U}(Y^{n},\phi^{n}_{s,\lambda})}\geq\lambda.
\label{ea1}
\end{equation}

\begin{remark}
Instead of (\ref{ea1}), the stronger utility hypothesis testing guarantee in (\ref{ed4}) is imposed here to make the following asymptotic analysis tractable.
\end{remark}

Let $\Phi^{n}_{s,\lambda}$ denote the set of all feasible $n$-slot policies satisfying the constraints in (\ref{eb1}) and (\ref{ed4}). In order to protect the privacy, an optimal management policy within $\Phi^{n}_{s,\lambda}$ is used to achieve the maximum minimal error probability of privacy hypothesis test as
\begin{equation}
\alpha_{P}^{*}(Y^{n},s,\lambda)=\max_{\phi_{s,\lambda}^{n}\in\Phi_{s,\lambda}^{n}}\alpha_{P}(Y^{n},\phi_{s,\lambda}^{n}),
\label{ed5}
\end{equation}
or equivalently the minimum error exponent $\frac{1}{n}\log\frac{1}{\alpha_{P}^{*}(Y^{n},s,\lambda)}$. The problem (\ref{ed5}) tradeoffs the hypothesis testing performances by minimizing the error exponent of the privacy hypothesis test and meanwhile guaranteeing a lower bound on the utility hypothesis testing performance.

The formulated problem (\ref{ed5}) corresponds to the practical scenario where the adversary is authorized and informed, e.g., a compromised grid operator in the smart meter privacy problem. In this case, the privacy leakage rate is usually more meaningful for the privacy measure when the adversary has a sequence of observations. In the following theorem, the asymptotic minimum error exponent of the privacy hypothesis test is characterized.

\begin{theorem}
Given feasible $s\geq0$ and $\lambda\geq0$, we have
\begin{equation}
\begin{aligned}
&\liminf_{n\to\infty}\frac{1}{n}\log\frac{1}{\alpha_{P}^{*}(Y^{n},s,\lambda)}\\
&\;\;\;=\inf_{n\in\mathbb{Z}_{+}}\min_{\phi_{s,\lambda}^{n}\in\Phi_{s,\lambda}^{n}}\frac{1}{n}\min_{\bar{u},\tilde{u}\in\{0,1\}}\left\{C(p_{Y^{n}|\bar{u},1}||p_{Y^{n}|\tilde{u},0})\right\}\\
&\;\;\;=\liminf_{n\to\infty}\min_{\phi_{s,\lambda}^{n}\in\Phi_{s,\lambda}^{n}}\frac{1}{n}\min_{\bar{u},\tilde{u}\in\{0,1\}}\left\{C(p_{Y^{n}|\bar{u},1}||p_{Y^{n}|\tilde{u},0})\right\}.
\end{aligned}
\label{ed7}
\end{equation}
\label{td1}
\end{theorem}

\begin{IEEEproof}
Given any $k\in\mathbb{Z}_{+}$, any $\phi_{s,\lambda}^{k}\in\Phi_{s,\lambda}^{k}$, the resulting pmfs $p_{Y^{k}|0,0}$, $p_{Y^{k}|0,1}$, $p_{Y^{k}|1,0}$, and $p_{Y^{k}|1,1}$, let $(\phi_{s,\lambda}^{k})^{l}$ denote a $kl$-slot policy which repeatedly uses $\phi_{s,\lambda}^{k}$ for $l$ times. The policy $(\phi_{s,\lambda}^{k})^{l}$ satisfies the constraint in (\ref{eb1}) over the $kl$-slot time horizon. Further, the resulting pmfs of the $kl$-slot policy $(\phi_{s,\lambda}^{k})^{l}$ jointly satisfy
\begin{equation*}
\begin{aligned}
\frac{1}{kl}&\min_{\bar{p},\tilde{p}\in\{0,1\}}\left\{C(p_{Y^{kl}|1,\bar{p}}||p_{Y^{kl}|0,\tilde{p}})\right\}\\
&=\frac{1}{k}\min_{\bar{p},\tilde{p}\in\{0,1\}}\left\{C(p_{Y^{k}|1,\bar{p}}||p_{Y^{k}|0,\tilde{p}})\right\}\\
&\geq\lambda+\frac{\log8p_{\max}}{k}
\geq\lambda+\frac{\log8p_{\max}}{kl}.
\end{aligned}
\end{equation*}
Therefore, we have $(\phi_{s,\lambda}^{k})^{l}\in\Phi_{s,\lambda}^{kl}$ since the policy $(\phi_{s,\lambda}^{k})^{l}$ also satisfies the constraint in (\ref{ed4}). It follows that
\begin{equation}
\begin{aligned}
&\liminf_{n\to\infty}\frac{1}{n}\log\frac{1}{\alpha_{P}^{*}(Y^{n},s,\lambda)}\\
&\;\;\;\leq\liminf_{l\to\infty}\frac{1}{kl}\log\frac{1}{\alpha_{P}^{*}(Y^{kl},s,\lambda)}\\
&\;\;\;\overset{(a)}{\leq}\lim_{l\to\infty}\frac{1}{kl}\log\frac{1}{\alpha_{P}(Y^{kl},(\phi_{s,\lambda}^{k})^{l})}\\
&\;\;\;\overset{(b)}{=}\frac{1}{k}\min_{\bar{u},\tilde{u}\in\{0,1\}}\left\{C(p_{Y^{k}|\bar{u},1}||p_{Y^{k}|\tilde{u},0})\right\},
\end{aligned}
\label{ed9}
\end{equation}
where the inequality $(a)$ follows from the optimization in the definition (\ref{ed5}); and the equality $(b)$ follows from Theorem \ref{tc1}. The inequality (\ref{ed9}) holds for all $k\in\mathbb{Z}_{+}$ and all $\phi_{s,\lambda}^{k}\in\Phi_{s,\lambda}^{k}$. Therefore, we have
\begin{equation}
\begin{aligned}
&\liminf_{n\to\infty}\frac{1}{n}\log\frac{1}{\alpha_{P}^{*}(Y^{n},s,\lambda)}\\
&\;\;\;\leq\inf_{k\in\mathbb{Z}_{+}}\min_{\phi_{s,\lambda}^{k}\in\Phi_{s,\lambda}^{k}}\frac{1}{k}\min_{\bar{u},\tilde{u}\in\{0,1\}}\left\{C(p_{Y^{k}|\bar{u},1}||p_{Y^{k}|\tilde{u},0})\right\}.
\end{aligned}
\label{ed20}
\end{equation}

Suppose that the policy $\phi_{s,\lambda}^{n*}\in\Phi_{s,\lambda}^{n}$ achieves $\alpha_{P}^{*}(Y^{n},s,\lambda)$, i.e.,
\begin{equation*}
\alpha_{P}^{*}(Y^{n},s,\lambda)=\alpha_{P}(Y^{n},\phi_{s,\lambda}^{n*}).
\end{equation*}
Let $p_{Y^{n}|0,0}^{*}$, $p_{Y^{n}|0,1}^{*}$, $p_{Y^{n}|1,0}^{*}$, and $p_{Y^{n}|1,1}^{*}$ denote the resulting pmfs. It follows from Proposition \ref{ld1} that
\begin{equation*}
\begin{aligned}
\frac{1}{n}&\log\frac{1}{\alpha_{P}^{*}(Y^{n},s,\lambda)}\\
&\geq\frac{1}{n}\min_{\bar{u},\tilde{u}\in\{0,1\}}\left\{C(p_{Y^{n}|\bar{u},1}^{*}||p_{Y^{n}|\tilde{u},0}^{*})\right\}-\frac{\log8p_{\max}}{n}\\
&\geq\min_{\phi_{s,\lambda}^{n}\in\Phi_{s,\lambda}^{n}}\frac{1}{n}\min_{\bar{u},\tilde{u}\in\{0,1\}}\left\{C(p_{Y^{n}|\bar{u},1}||p_{Y^{n}|\tilde{u},0})\right\}-\frac{\log8p_{\max}}{n}.
\end{aligned}
\end{equation*}
In the asymptotic regime as $n\to\infty$, we have the following lower bound:
\begin{equation}
\begin{aligned}
&\liminf_{n\to\infty}\frac{1}{n}\log\frac{1}{\alpha_{P}^{*}(Y^{n},s,\lambda)}\\
&\;\;\;\geq\liminf_{n\to\infty}\min_{\phi_{s,\lambda}^{n}\in\Phi_{s,\lambda}^{n}}\frac{1}{n}\min_{\bar{u},\tilde{u}\in\{0,1\}}\left\{C(p_{Y^{n}|\bar{u},1}||p_{Y^{n}|\tilde{u},0})\right\}.
\end{aligned}
\label{ed21}
\end{equation}

From the definitions of infimum and limit infimum, we have
\begin{equation}
\begin{aligned}
&\inf_{k\in\mathbb{Z}_{+}}\min_{\phi_{s,\lambda}^{k}\in\Phi_{s,\lambda}^{k}}\frac{1}{k}\min_{\bar{u},\tilde{u}\in\{0,1\}}\left\{C(p_{Y^{k}|\bar{u},1}||p_{Y^{k}|\tilde{u},0})\right\}\\
&\;\;\;\leq\liminf_{n\to\infty}\min_{\phi_{s,\lambda}^{n}\in\Phi_{s,\lambda}^{n}}\frac{1}{n}\min_{\bar{u},\tilde{u}\in\{0,1\}}\left\{C(p_{Y^{n}|\bar{u},1}||p_{Y^{n}|\tilde{u},0})\right\}.
\end{aligned}
\label{ed25}
\end{equation}

The inequalities (\ref{ed20}), (\ref{ed21}), and (\ref{ed25}) jointly lead to the asymptotic minimum error exponent in (\ref{ed7}).
\end{IEEEproof}

\begin{remark}
The second equality in Theorem \ref{td1}
can be alternatively justified by the following inequality: For all $k$, $l\in\mathbb{Z}_{+}$, we have
\begin{equation}
\begin{aligned}
&\min_{\phi_{s,\lambda}^{k}\in\Phi_{s,\lambda}^{k}}\frac{1}{k}\min_{\bar{u},\tilde{u}\in\{0,1\}}\left\{C(p_{Y^{k}|\bar{u},1}||p_{Y^{k}|\tilde{u},0})\right\}\\
&\;\;\;\geq\min_{\phi_{s,\lambda}^{n}\in\Phi_{s,\lambda}^{n}}\frac{1}{n}\min_{\bar{u},\tilde{u}\in\{0,1\}}\left\{C(p_{Y^{n}|\bar{u},1}||p_{Y^{n}|\tilde{u},0})\right\},
\end{aligned}
\label{ed16}
\end{equation}
where $n=kl\geq k$.
\label{rd1}
\end{remark}

The proof ideas of the inequality (\ref{ed16}) are: Construct a $kl$-slot policy $(\phi_{s,\lambda}^{k\star})^{l}$ where $\phi_{s,\lambda}^{k\star}$ achieves the minimal Chernoff information rate over policies in $\Phi_{s,\lambda}^{k}$; show that the constructed policy $(\phi_{s,\lambda}^{k\star})^{l}$ achieves the same minimal Chernoff information rate as $\phi_{s,\lambda}^{k\star}$; and prove the inequality by the fact that $(\phi_{s,\lambda}^{k\star})^{l}$ does not necessarily achieve the minimal Chernoff information rate over policies in $\Phi_{s,\lambda}^{n}$.

Theorem \ref{td1} shows that the asymptotic minimum error exponent of the privacy hypothesis test is the infimum of corresponding minimal Chernoff information rates subject to the utility hypothesis testing guarantees. Theorem \ref{td1} also shows that the infimum of minimal Chernoff information rates in the general case is taken at the limit of the block length $n\to\infty$. Therefore, the numerical evaluation of the asymptotic minimum error exponent of the privacy hypothesis test and the design of an asymptotically optimal management policy are difficult tasks.

For all $k\in\mathbb{Z}_{+}$, the minimal Chernoff information rate $\min_{\phi_{s,\lambda}^{k}\in\Phi_{s,\lambda}^{k}}\frac{1}{k}\min_{\bar{u},\tilde{u}\in\{0,1\}}\left\{C(p_{Y^{k}|\bar{u},1}||p_{Y^{k}|\tilde{u},0})\right\}$ is the asymptotic error exponent of the privacy hypothesis test when the block-wise i.i.d. management policy $(\phi_{s,\lambda}^{k\star})^{l}$ is used with
\begin{equation*}
\phi_{s,\lambda}^{k\star}=\argmin_{\phi_{s,\lambda}^{k}\in\Phi_{s,\lambda}^{k}}\frac{1}{k}\min_{\bar{u},\tilde{u}\in\{0,1\}}\left\{C(p_{Y^{k}|\bar{u},1}||p_{Y^{k}|\tilde{u},0})\right\},
\end{equation*}
i.e.,
\begin{equation*}
\begin{aligned}
&\lim_{l\to\infty}\frac{1}{kl}\log\frac{1}{\alpha_{P}(Y^{kl},(\phi_{s,\lambda}^{k\star})^{l})}\\
&\;\;\;=\min_{\phi_{s,\lambda}^{k}\in\Phi_{s,\lambda}^{k}}\frac{1}{k}\min_{\bar{u},\tilde{u}\in\{0,1\}}\left\{C(p_{Y^{k}|\bar{u},1}||p_{Y^{k}|\tilde{u},0})\right\}.
\end{aligned}
\end{equation*}
On the other hand, the minimal Chernoff information rate $\min_{\phi_{s,\lambda}^{k}\in\Phi_{s,\lambda}^{k}}\frac{1}{k}\min_{\bar{u},\tilde{u}\in\{0,1\}}\left\{C(p_{Y^{k}|\bar{u},1}||p_{Y^{k}|\tilde{u},0})\right\}$ is an upper bound on the asymptotic minimum error exponent of the privacy hypothesis test and therefore can be seen as an asymptotic privacy guarantee. Then, the evaluation of an asymptotic privacy guarantee and the design of the corresponding block-wise i.i.d. policy are tractable.

\section{Numerical Example}
\label{section5}
\begin{figure}
\centering
\includegraphics[scale=0.5]{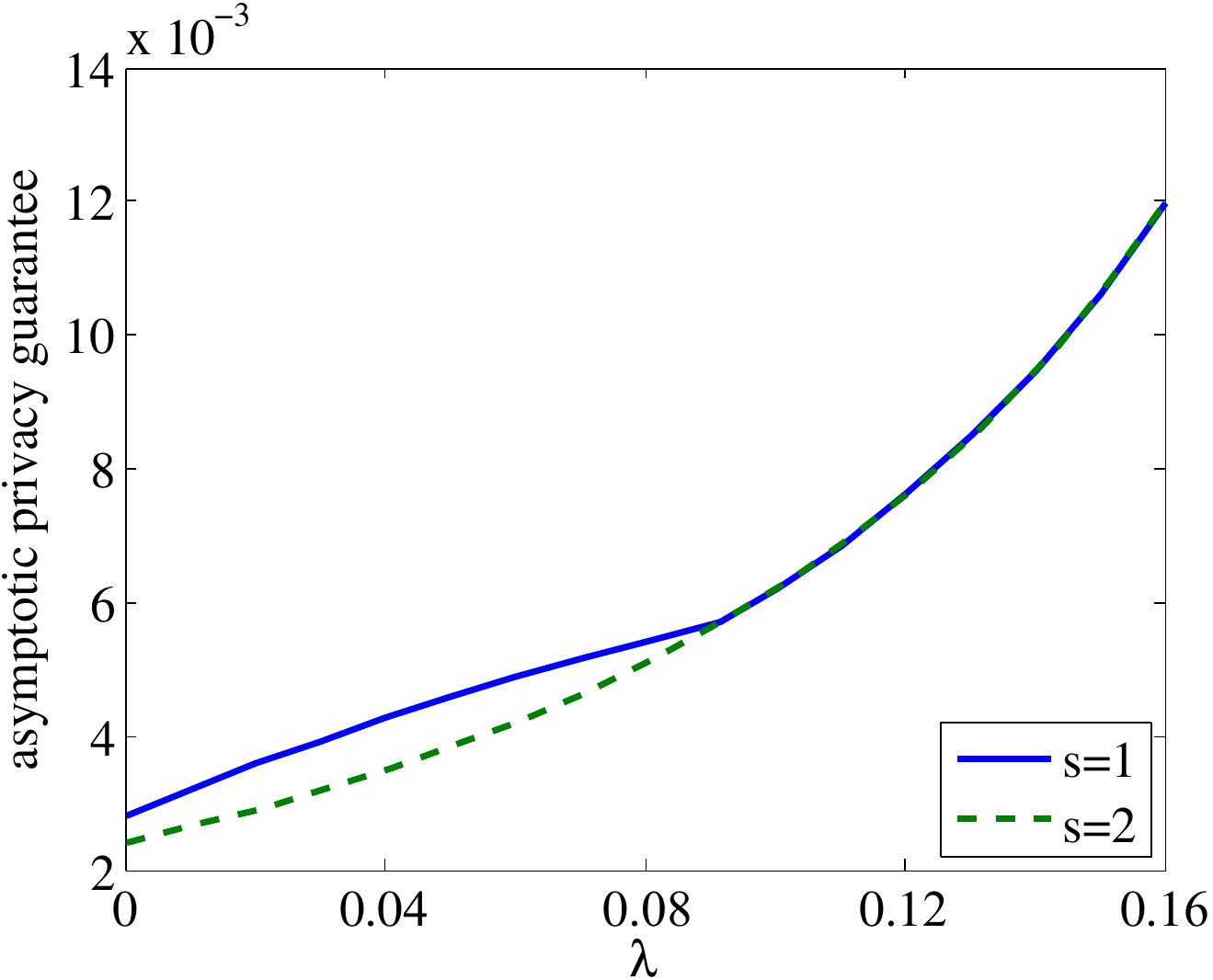}
\caption{Illustration of the trade-off between the asymptotic privacy guarantee $\min_{\phi_{s,\lambda}^{1}\in\Phi_{s,\lambda}^{1}}\min_{\bar{u},\tilde{u}\in\{0,1\}}\left\{C(p_{Y|\bar{u},1}||p_{Y|\tilde{u},0})\right\}$ and the utility hypothesis testing guarantee $\lambda$.}
\label{figure3}
\end{figure}

Fig. \ref{figure3} illustrates the trade-off of the asymptotic privacy guarantee $\min_{\phi_{s,\lambda}^{1}\in\Phi_{s,\lambda}^{1}}\min_{\bar{u},\tilde{u}\in\{0,1\}}\left\{C(p_{Y|\bar{u},1}||p_{Y|\tilde{u},0})\right\}$ and the utility hypothesis testing guarantee $\lambda$ in a simple model with $\mathcal{X}=\mathcal{Z}=\{0,1\}$. The parameters are set as: $0\leq\lambda\leq0.16$, $s=1,2$, $p_{\max}=\frac{1}{4}$, $p_{X|0,0}(0)=0.1$, $p_{X|0,1}(0)=0.25$, $p_{X|1,0}(0)=0.8$, $p_{X|1,1}(0)=0.9$, $p_{Z}(0)=0.2$. As expected, the value of the asymptotic privacy guarantee increases as the value of the utility hypothesis testing guarantee increases; and a greater value of $s$ leads to a better asymptotic privacy guarantee. We can also learn from Fig. \ref{figure3} that an asymptotic privacy guarantee is not necessarily convex or concave of the utility hypothesis testing guarantee $\lambda$. Since the asymptotic minimum error exponent of the privacy hypothesis test is the infimum of the asymptotic privacy guarantees, its convexity property is not clear.

\section{Conclusion}
\label{section6}
We showed that the error exponent of a Bayesian composite hypothesis test can be characterized by the corresponding minimal Chernoff information rate. With the optimal management, we further proved that the asymptotic minimum error exponent of the privacy hypothesis test can be characterized by the infimum of the corresponding minimal Chernoff information rates subject to utility hypothesis testing guarantees. The studied optimal management with privacy-utility trade-off can be applied in many practical scenarios, e.g., smart metering system.

\appendix
\label{app}
\subsection{Proof of Proposition \ref{lc1}}
Given a sequence $x^{n}$, an optimal hypothesis testing decision $\hat{U}$ is made according to the following strategy:
\begin{equation}
\frac{\sum_{p=0}^{1}p_{X^{n}|0,p}(x^{n})p_{U,P}(0,p)}{\sum_{p=0}^{1}p_{X^{n}|1,p}(x^{n})p_{U,P}(1,p)}\overset{\hat{U}=0}{\underset{\hat{U}=1}{\gtrless}}1.
\label{ec1}
\end{equation}
Denote the type of the sequence $x^{n}$ by $t_{X}$. It follows from the {\it method of type} \cite{csiszar1998} that
\begin{equation}
p_{X^{n}|u,p}(x^{n})=\exp(-nD(t_{X}||p_{X|u,p})-nH(t_{X})),
\label{ec2}
\end{equation}
i.e., sequences of the same type have the same probability under a hypothesis pair realization $(u,p)\in\mathcal{U}\times\mathcal{P}$. The optimal hypothesis testing strategy (\ref{ec1}) can be equivalently reformulated as
\begin{equation}
\frac{\sum_{p=0}^{1}\exp(-nD(t_{X}||p_{X|0,p}))p_{U,P}(0,p)}{\sum_{p=0}^{1}\exp(-nD(t_{X}||p_{X|1,p}))p_{U,P}(1,p)}\overset{\hat{U}=0}{\underset{\hat{U}=1}{\gtrless}}1.
\label{ec3}
\end{equation}

\begin{proposition}
In the asymptotic regime as $n\to\infty$, the optimal hypothesis testing strategy (\ref{ec1}) reduces to the following test depending on the type $t_{X}$ only:
\begin{equation}
\frac{\min_{p\in\{0,1\}}\{D(t_{X}||p_{X|0,p})\}}{\min_{p\in\{0,1\}}\{D(t_{X}||p_{X|1,p})\}}\overset{\hat{U}=1}{\underset{\hat{U}=0}{\gtrless}}1.
\label{ec4}
\end{equation}
\label{pc1}
\end{proposition}

\begin{IEEEproof}
W.l.o.g., we consider a type $t_{X}$ such that
\begin{equation*}
\frac{\min_{p\in\{0,1\}}\{D(t_{X}||p_{X|0,p})\}}{\min_{p\in\{0,1\}}\{D(t_{X}||p_{X|1,p})\}}<1.
\end{equation*}
The test ratio in the optimal strategy (\ref{ec3}) can be rewritten as
\begin{equation*}
\frac{\sum_{p'=0}^{1}\frac{\exp(-nD(t_{X}||p_{X|0,p'}))}{\exp(-n\min_{p\in\{0,1\}}\{D(t_{X}||p_{X|0,p})\})}p_{U,P}(0,p')}{\sum_{p'=0}^{1}\frac{\exp(-nD(t_{X}||p_{X|1,p'}))}{\exp(-n\min_{p\in\{0,1\}}\{D(t_{X}||p_{X|0,p})\})}p_{U,P}(1,p')}.
\end{equation*}
As $n\to\infty$, we have
\begin{equation*}
\frac{\sum_{p'=0}^{1}\frac{\exp(-nD(t_{X}||p_{X|0,p'}))}{\exp(-n\min_{p\in\{0,1\}}\{D(t_{X}||p_{X|0,p})\})}p_{U,P}(0,p')}{\sum_{p'=0}^{1}\frac{\exp(-nD(t_{X}||p_{X|1,p'}))}{\exp(-n\min_{p\in\{0,1\}}\{D(t_{X}||p_{X|0,p})\})}p_{U,P}(1,p')}>1,
\end{equation*}
i.e., the hypothesis testing decision is $\hat{U}=0$.
\end{IEEEproof}

\begin{IEEEproof}[Proof of Proposition \ref{lc1}]
We define two type sets as
\begin{equation*}
\begin{aligned}
\mathcal{A}_{\hat{U}=0}=&\,\left\{t_{X}:\frac{\min_{p\in\{0,1\}}\{D(t_{X}||p_{X|0,p})\}}{\min_{p\in\{0,1\}}\{D(t_{X}||p_{X|1,p})\}}\leq1\right\},\\
\mathcal{A}_{\hat{U}=1}=&\,\left\{t_{X}:\frac{\min_{p\in\{0,1\}}\{D(t_{X}||p_{X|0,p})\}}{\min_{p\in\{0,1\}}\{D(t_{X}||p_{X|1,p})\}}\geq1\right\}.
\end{aligned}
\end{equation*}
When $n\to\infty$ and the optimal hypothesis testing strategy (\ref{ec4}) is used, we have
\begin{equation*}
\begin{aligned}
\alpha_{U}(X^{n})
\doteq&\,\exp(-n\cdot\min_{t_{X}\in\mathcal{A}_{\hat{U}=1}}D(t_{X}||p_{X|0,0}))p_{U,P}(0,0)\\
&\,+\exp(-n\cdot\min_{t_{X}\in\mathcal{A}_{\hat{U}=1}}D(t_{X}||p_{X|0,1}))p_{U,P}(0,1)\\
&\,+\exp(-n\cdot\min_{t_{X}\in\mathcal{A}_{\hat{U}=0}}D(t_{X}||p_{X|1,0}))p_{U,P}(1,0)\\
&\,+\exp(-n\cdot\min_{t_{X}\in\mathcal{A}_{\hat{U}=0}}D(t_{X}||p_{X|1,1}))p_{U,P}(1,1)\\
\doteq&\,\exp\left(-n\cdot\min\left\{\begin{gathered}
\min_{t_{X}\in\mathcal{A}_{\hat{U}=1}}D(t_{X}||p_{X|0,0})\\
\min_{t_{X}\in\mathcal{A}_{\hat{U}=1}}D(t_{X}||p_{X|0,1})\\
\min_{t_{X}\in\mathcal{A}_{\hat{U}=0}}D(t_{X}||p_{X|1,0})\\
\min_{t_{X}\in\mathcal{A}_{\hat{U}=0}}D(t_{X}||p_{X|1,1})
\end{gathered}\right\}\right),
\end{aligned}
\end{equation*}
where $\doteq$ means equality to the first order in the exponent as defined in \cite[(3.26)]{cover2006}; the first approximation follows from Sanov's theorem; and the second follows since the exponential rate is determined by the minimal exponent. Therefore, the asymptotic error exponent of utility hypothesis test with the i.i.d. sequence $X^{n}$ is
\begin{equation}
\lim_{n\to\infty}\frac{1}{n}\log\frac{1}{\alpha_{U}(X^{n})}
=\min\left\{\begin{gathered}
\min_{t_{X}\in\mathcal{A}_{\hat{U}=1}}D(t_{X}||p_{X|0,0})\\
\min_{t_{X}\in\mathcal{A}_{\hat{U}=1}}D(t_{X}||p_{X|0,1})\\
\min_{t_{X}\in\mathcal{A}_{\hat{U}=0}}D(t_{X}||p_{X|1,0})\\
\min_{t_{X}\in\mathcal{A}_{\hat{U}=0}}D(t_{X}||p_{X|1,1})
\end{gathered}\right\}.
\label{ec5}
\end{equation}
Note that a Kullback-Leibler divergence $D(t_{X}||p_{X|u,p})$ is a convex function of $t_{X}$; an optimization domain $\mathcal{A}_{\hat{U}=\hat{u}}$ is generally not a convex set but a union of two convex sets confined by hyperplanes\footnote{
\begin{equation*}
\begin{aligned}
\mathcal{A}_{\hat{U}=0}=&\,\bigg\{t_{X}:\underbrace{\sum_{x\in\mathcal{X}}t_{X}(x)\log\frac{p_{X|1,p}(x)}{p_{X|0,0}(x)}\leq0}_{D(t_{X}||p_{X|0,0})\leq D(t_{X}||p_{X|1,p})},\forall p\in\mathcal{P}\bigg\}\\
&\,\cup\bigg\{t_{X}:\underbrace{\sum_{x\in\mathcal{X}}t_{X}(x)\log\frac{p_{X|1,p}(x)}{p_{X|0,1}(x)}\leq0}_{D(t_{X}||p_{X|0,1})\leq D(t_{X}||p_{X|1,p})},\forall p\in\mathcal{P}\bigg\},
\end{aligned}
\end{equation*}
\begin{equation*}
\begin{aligned}
\mathcal{A}_{\hat{U}=1}=&\,\bigg\{t_{X}:\underbrace{\sum_{x\in\mathcal{X}}t_{X}(x)\log\frac{p_{X|0,p}(x)}{p_{X|1,0}(x)}\leq0}_{D(t_{X}||p_{X|1,0})\leq D(t_{X}||p_{X|0,p})},\forall p\in\mathcal{P}\bigg\}\\
&\,\cup\bigg\{t_{X}:\underbrace{\sum_{x\in\mathcal{X}}t_{X}(x)\log\frac{p_{X|0,p}(x)}{p_{X|1,1}(x)}\leq0}_{D(t_{X}||p_{X|1,1})\leq D(t_{X}||p_{X|0,p})},\forall p\in\mathcal{P}\bigg\}.
\end{aligned}
\end{equation*}};
and furthermore $p_{X|1,p}\not\in\mathcal{A}_{\hat{U}=0}$, $p_{X|0,p}\not\in\mathcal{A}_{\hat{U}=1}$, for all $p\in\{0,1\}$.

The minimizations on the right-hand side of (\ref{ec5}) can be specified as
\begin{equation*}
\begin{aligned}
&\lim_{n\to\infty}\frac{1}{n}\log\frac{1}{\alpha_{U}(X^{n})}\\
&\;\;\;=\min\left\{\begin{gathered}
\begin{aligned}
\min_{t_{X}}&\,\min\{D(t_{X}||p_{X|0,0}),D(t_{X}||p_{X|0,1})\}\\
\textnormal{s.t.}&\,D(t_{X}||p_{X|1,0})\leq D(t_{X}||p_{X|0,0})\\
&\,D(t_{X}||p_{X|1,0})\leq D(t_{X}||p_{X|0,1})
\end{aligned}\\
\begin{aligned}
\min_{t_{X}}&\,\min\{D(t_{X}||p_{X|0,0}),D(t_{X}||p_{X|0,1})\}\\
\textnormal{s.t.}&\,D(t_{X}||p_{X|1,1})\leq D(t_{X}||p_{X|0,0})\\
&\,D(t_{X}||p_{X|1,1})\leq D(t_{X}||p_{X|0,1})
\end{aligned}\\
\begin{aligned}
\min_{t_{X}}&\,\min\{D(t_{X}||p_{X|1,0}),D(t_{X}||p_{X|1,1})\}\\
\textnormal{s.t.}&\,D(t_{X}||p_{X|0,0})\leq D(t_{X}||p_{X|1,0})\\
&\,D(t_{X}||p_{X|0,0})\leq D(t_{X}||p_{X|1,1})
\end{aligned}\\
\begin{aligned}
\min_{t_{X}}&\,\min\{D(t_{X}||p_{X|1,0}),D(t_{X}||p_{X|1,1})\}\\
\textnormal{s.t.}&\,D(t_{X}||p_{X|0,1})\leq D(t_{X}||p_{X|1,0})\\
&\,D(t_{X}||p_{X|0,1})\leq D(t_{X}||p_{X|1,1})
\end{aligned}
\end{gathered}\right\}.
\end{aligned}
\end{equation*}
There are eight inner minimizations. Due to the symmetric formulation, we focus on the first inner minimization, which consists of a convex objective and two affine inequality constraints, i.e., it is a convex optimization and satisfies the Slater's condition \cite{boyd2004} for strong duality. It follows that
\begin{equation*}
\begin{aligned}
&\begin{aligned}
\min_{t_{X}}&\,D(t_{X}||p_{X|0,0})\\
\textnormal{s.t.}&\,D(t_{X}||p_{X|1,0})\leq D(t_{X}||p_{X|0,0})\\
&\,D(t_{X}||p_{X|1,0})\leq D(t_{X}||p_{X|0,1})
\end{aligned}\\
&\;\;\;=\max_{\mu\geq0,\nu\geq0}-\log\sum_{x\in\mathcal{X}}p_{X|1,0}^{\mu+\nu}(x)p_{X|0,0}^{1-\mu}(x)p_{X|0,1}^{-\nu}(x),
\end{aligned}
\end{equation*}
where the dual objective function is jointly concave of the dual variables $\mu$ and $\nu$. Given $\mu'=1$ and $\nu'=0$, we have the following lower bound:
\begin{equation*}
\begin{aligned}
&\max_{\mu\geq0,\nu\geq0}-\log\sum_{x\in\mathcal{X}}p_{X|1,0}^{\mu+\nu}(x)p_{X|0,0}^{1-\mu}(x)p_{X|0,1}^{-\nu}(x)\\
&\;\;\;\geq-\log\sum_{x\in\mathcal{X}}p_{X|1,0}^{\mu'+\nu'}(x)p_{X|0,0}^{1-\mu'}(x)p_{X|0,1}^{-\nu'}(x)=0.
\end{aligned}
\end{equation*}
For all $\mu\geq0$ and $\nu\geq0$, we have the following upper bound from Jensen's inequality \cite[Theorem 2.6.2]{cover2006}:
\begin{equation}
\begin{aligned}
&-\log\sum_{x\in\mathcal{X}}p_{X|1,0}^{\mu+\nu}(x)p_{X|0,0}^{1-\mu}(x)p_{X|0,1}^{-\nu}(x)\\
&\;\;\;=-\log\sum_{x\in\mathcal{X}}p_{X|1,0}(x)\left(\frac{p_{X|1,0}(x)}{p_{X|0,0}(x)}\right)^{\mu-1}\left(\frac{p_{X|1,0}(x)}{p_{X|0,1}(x)}\right)^{\nu}\\
&\;\;\;\leq-\sum_{x\in\mathcal{X}}p_{X|1,0}(x)\log\left(\left(\frac{p_{X|1,0}(x)}{p_{X|0,0}(x)}\right)^{\mu-1}\left(\frac{p_{X|1,0}(x)}{p_{X|0,1}(x)}\right)^{\nu}\right)\\
&\;\;\;=(1-\mu)D(p_{X|1,0}||p_{X|0,0})-\nu D(p_{X|1,0}||p_{X|0,1}).
\end{aligned}
\label{ea20}
\end{equation}
For all $\mu'\geq1$ and $\nu'\geq0$, it follows from the non-negativity of Kullback-Leibler divergence that
\begin{equation*}
\begin{aligned}
&-\log\sum_{x\in\mathcal{X}}p_{X|1,0}^{\mu'+\nu'}(x)p_{X|0,0}^{1-\mu'}(x)p_{X|0,1}^{-\nu'}(x)\\
&\;\;\;\leq(1-\mu')D(p_{X|1,0}||p_{X|0,0})-\nu' D(p_{X|1,0}||p_{X|0,1})\leq0.
\end{aligned}
\end{equation*}
Note the zero lower bound. It is sufficient to consider the dual variable $\mu$ within the interval $[0,1]$, i.e.,
\begin{equation*}
\begin{aligned}
&\max_{\mu\geq0,\nu\geq0}-\log\sum_{x\in\mathcal{X}}p_{X|1,0}^{\mu+\nu}(x)p_{X|0,0}^{1-\mu}(x)p_{X|0,1}^{-\nu}(x)\\
&\;\;\;=\max_{1\geq\mu\geq0,\nu\geq0}-\log\sum_{x\in\mathcal{X}}p_{X|1,0}^{\mu+\nu}(x)p_{X|0,0}^{1-\mu}(x)p_{X|0,1}^{-\nu}(x)\\
&\;\;\;=\max_{1\geq\mu\geq0,\nu\geq0}T_{\mu,\nu}(p_{X|1,0}||p_{X|0,0};p_{X|0,1})\\
&\;\;\;=T(p_{X|1,0}||p_{X|0,0};p_{X|0,1}).
\end{aligned}
\end{equation*}
\end{IEEEproof}

\subsection{Proof of Lemma \ref{lc2}}
From the proof of Proposition \ref{lc1}, we can summarize the following properties of functions $T$ and $T_{\mu,\nu}$.

\begin{proposition}
Given pmfs $Q_{1}$, $Q_{2}$, and $Q_{3}$, the function $T_{\mu,\nu}(Q_{1}||Q_{2};Q_{3})$ is jointly concave of the dual variables $\mu$ and $\nu$; and the function $T(Q_{1}||Q_{2};Q_{3})$ is non-negative.
\label{pc3}
\end{proposition}


\begin{IEEEproof}[Proof of Lemma \ref{lc2}]
From the inequality in (\ref{ea20}), the non-negativity of Kullback-Leibler divergence, and the non-negativity of $T(Q_{1}||Q_{2};Q_{3})$, we can impose a stronger constraint on the optimization region of $(\mu,\nu)$ in the definition of $T(Q_{1}||Q_{2};Q_{3})$ as
\begin{equation*}
\mathcal{B}_{1}=\left\{(\mu,\nu):\mu\geq0,\nu\geq0,(1-\mu)\frac{D(Q_{1}||Q_{2})}{D(Q_{1}||Q_{3})}\geq\nu\right\},
\end{equation*}
which is a triangular region with corners $(0,0)$, $(1,0)$, and $\left(0,\frac{D(Q_{1}||Q_{2})}{D(Q_{1}||Q_{3})}\right)$, i.e.,
\begin{equation*}
\begin{aligned}
&T(Q_{1}||Q_{2};Q_{3})\\
&\;\;\;=\max_{(\mu,\nu)\in\mathcal{B}_{1}}-\log\sum_{a\in\mathcal{S}}Q_{1}^{\mu+\nu}(a)Q_{2}^{1-\mu}(a)Q_{3}^{-\nu}(a).
\end{aligned}
\end{equation*}
Similarly, we have
\begin{equation*}
\begin{aligned}
&T(Q_{1}||Q_{3};Q_{2})\\
&\;\;\;=\max_{(\mu',\nu')\in\mathcal{B}'_{2}}-\log\sum_{a\in\mathcal{S}}Q_{1}^{\mu'+\nu'}(a)Q_{2}^{-\nu'}(a)Q_{3}^{1-\mu'}(a),
\end{aligned}
\end{equation*}
where
\begin{equation*}
\mathcal{B}'_{2}=\left\{(\mu',\nu'):\mu'\geq0,\nu'\geq0,(1-\mu')\frac{D(Q_{1}||Q_{3})}{D(Q_{1}||Q_{2})}\geq\nu'\right\}.
\end{equation*}
Let $-\nu'=1-\mu$ and $1-\mu'=-\nu$. Then, we have
\begin{equation*}
\begin{aligned}
&T(Q_{1}||Q_{3};Q_{2})\\
&\;\;\;=\max_{(\mu,\nu)\in\mathcal{B}_{2}}-\log\sum_{a\in\mathcal{S}}Q_{1}^{\mu+\nu}(a)Q_{2}^{1-\mu}(a)Q_{3}^{-\nu}(a),
\end{aligned}
\end{equation*}
where
\begin{equation*}
\mathcal{B}_{2}=\left\{(\mu,\nu):\mu\geq1,\nu\geq-1,(1-\mu)\frac{D(Q_{1}||Q_{2})}{D(Q_{1}||Q_{3})}\geq\nu\right\}
\end{equation*}
is a triangular region with corners $(1,0)$, $(1,-1)$, and $\left(1+\frac{D(Q_{1}||Q_{3})}{D(Q_{1}||Q_{2})},-1\right)$.
Now the original problem reduces to an optimization of the objective $T_{\mu,\nu}(Q_{1}||Q_{2};Q_{3})$ over the two triangular regions $\mathcal{B}_{1}$ and $\mathcal{B}_{2}$ as illustrated in Fig. \ref{figure2}:
\begin{equation*}
\min\left\{\begin{gathered}
T(Q_{1}||Q_{2};Q_{3})\\
T(Q_{1}||Q_{3};Q_{2})
\end{gathered}\right\}
=\min\left\{\begin{gathered}
\max_{(\mu,\nu)\in\mathcal{B}_{1}}T_{\mu,\nu}(Q_{1}||Q_{2};Q_{3})\\
\max_{(\mu,\nu)\in\mathcal{B}_{2}}T_{\mu,\nu}(Q_{1}||Q_{2};Q_{3})
\end{gathered}\right\}.
\end{equation*}

\begin{figure}
\centering
\includegraphics[scale=1]{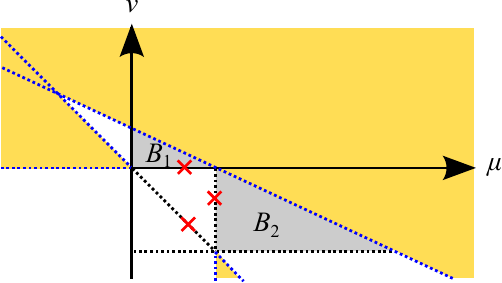}
\caption{Illustration of optimization regions $\mathcal{B}_{1}$ and $\mathcal{B}_{2}$, where $0<\frac{D(Q_{1}||Q_{2})}{D(Q_{1}||Q_{3})}<1$; any point on the blue dotted lines leads to a non-positive value of $T_{\mu,\nu}(Q_{1}||Q_{2};Q_{3})$; the red crosses on the lines $\nu=0$, $\mu=1$, and $\nu=-\mu$ lead to non-negative values of $T_{\mu,\nu}(Q_{1}||Q_{2};Q_{3})$ as Chernoff informations $C(Q_{1}||Q_{2})$, $C(Q_{1}||Q_{3})$, and $C(Q_{2}||Q_{3})$, respectively.}
\label{figure2}
\end{figure}

Given pmfs $Q_{1}$, $Q_{2}$, and $Q_{3}$, we can identify more properties of the objective function $T_{\mu,\nu}(Q_{1}||Q_{2};Q_{3})$ over the space $(\mu,\nu)\in\mathbb{R}^{2}$.

When $\nu=0$, $T_{\mu,\nu}(Q_{1}||Q_{2};Q_{3})$ reduces to a function of $\mu$ as $-\log\sum_{a\in\mathcal{S}}Q_{1}^{\mu}(a)Q_{2}^{1-\mu}(a)$. Note that $-\log\sum_{a\in\mathcal{S}}Q_{1}^{0}(a)Q_{2}^{1}(a)=-\log\sum_{a\in\mathcal{S}}Q_{1}^{1}(a)Q_{2}^{0}(a)=0$; $C(Q_{1}||Q_{2})=\max_{1\geq\mu\geq0}-\log\sum_{a\in\mathcal{S}}Q_{1}^{\mu}(a)Q_{2}^{1-\mu}(a)\geq0$; and the concavity of $-\log\sum_{a\in\mathcal{S}}Q_{1}^{\mu}(a)Q_{2}^{1-\mu}(a)$ over $\mu\in\mathbb{R}$. We can conclude that $T_{\mu,\nu}(Q_{1}||Q_{2};Q_{3})\leq0$ when $\mu\leq0$ and $\nu=0$.

When $\mu=1$, $T_{\mu,\nu}(Q_{1}||Q_{2};Q_{3})$ reduces to a function of $\nu$ as $-\log\sum_{a\in\mathcal{S}}Q_{1}^{1+\nu}(a)Q_{3}^{-\nu}(a)$. Note that $-\log\sum_{a\in\mathcal{S}}Q_{1}^{0}(a)Q_{3}^{1}(a)=-\log\sum_{a\in\mathcal{S}}Q_{1}^{1}(a)Q_{3}^{0}(a)=0$; $C(Q_{1}||Q_{3})=\max_{0\geq\nu\geq-1}-\log\sum_{a\in\mathcal{S}}Q_{1}^{1+\nu}(a)Q_{3}^{-\nu}(a)\geq0$; and the concavity of $-\log\sum_{a\in\mathcal{S}}Q_{1}^{1+\nu}(a)Q_{3}^{-\nu}(a)$ over $\nu\in\mathbb{R}$. We can conclude that $T_{\mu,\nu}(Q_{1}||Q_{2};Q_{3})\leq0$ when $\mu=1$ and $\nu\leq-1$.

When $\mu=-\nu$, $T_{\mu,\nu}(Q_{1}||Q_{2};Q_{3})$ reduces to a function of $\mu$ as $-\log\sum_{a\in\mathcal{S}}Q_{2}^{1-\mu}(a)Q_{3}^{\mu}(a)$. Note that $-\log\sum_{a\in\mathcal{S}}Q_{2}^{0}(a)Q_{3}^{1}(a)=-\log\sum_{a\in\mathcal{S}}Q_{2}^{1}(a)Q_{3}^{0}(a)=0$; $C(Q_{2}||Q_{3})=\max_{1\geq\mu\geq0}-\log\sum_{a\in\mathcal{S}}Q_{2}^{1-\mu}(a)Q_{3}^{\mu}(a)\geq0$; and the concavity of $-\log\sum_{a\in\mathcal{S}}Q_{2}^{1-\mu}(a)Q_{3}^{\mu}(a)$ over $\mu\in\mathbb{R}$. We can conclude that $T_{\mu,\nu}(Q_{1}||Q_{2};Q_{3})\leq0$ when $\mu=-\nu\leq0$ and $\mu=-\nu\geq1$.

Jointly with the observation that $T_{\mu,\nu}(Q_{1}||Q_{2};Q_{3})\leq0$ when $(1-\mu)\frac{D(Q_{1}||Q_{2})}{D(Q_{1}||Q_{3})}=\nu$, we can conclude that the optimizer $(\mu^{*},\nu^{*})\in\mathbb{R}^{2}$ of the optimization $\max_{(\mu,\nu)\in\mathbb{R}^{2}}T_{\mu,\nu}(Q_{1}||Q_{2};Q_{3})$ cannot be within the yellow region as shown in Fig. \ref{figure2}. It is because that the assumption of $(\mu^{*},\nu^{*})$ in the yellow region will lead to non-positive values of $T_{\mu,\nu}(Q_{1}||Q_{2};Q_{3})$ to be achieved by all points in the two triangular regions $\mathcal{B}_{1}$ and $\mathcal{B}_{2}$.

No matter where the optimizer $(\mu^{*},\nu^{*})$ of the optimization $\max_{(\mu,\nu)\in\mathbb{R}^{2}}T_{\mu,\nu}(Q_{1}||Q_{2};Q_{3})$ locates in the remaining feasible region, the joint concavity of $T_{\mu,\nu}(Q_{1}||Q_{2};Q_{3})$ and these identified non-positive line segments will lead to the following equality
\begin{equation*}
\begin{aligned}
&\min\left\{\begin{gathered}
\max_{(\mu,\nu)\in\mathcal{B}_{1}}T_{\mu,\nu}(Q_{1}||Q_{2};Q_{3})\\
\max_{(\mu,\nu)\in\mathcal{B}_{2}}T_{\mu,\nu}(Q_{1}||Q_{2};Q_{3})
\end{gathered}\right\}\\
&\;\;\;=\min\left\{\begin{gathered}
\max_{1\geq\mu\geq0}-\log\sum_{a\in\mathcal{S}}Q_{1}^{\mu}(a)Q_{2}^{1-\mu}(a)\\
\max_{0\geq\nu\geq-1}-\log\sum_{a\in\mathcal{S}}Q_{1}^{1+\nu}(a)Q_{3}^{-\nu}(a)
\end{gathered}\right\}\\
&\;\;\;=\min\left\{\begin{gathered}
C(Q_{1}||Q_{2})\\
C(Q_{1}||Q_{3})
\end{gathered}\right\}.
\end{aligned}
\end{equation*}
The justification is trivial. Here, we only show the case of the optimizer $(\mu^{*},\nu^{*})$ locating within the triangular region $\mathcal{B}_{1}$. In this case, there always exist an intersection of the line segment $\nu=0$ with $1\geq\mu\geq0$ and the line segment between the optimizer $(\mu^{*},\nu^{*})$ and any point on the line segment $\mu=1$ with $0\geq\nu\geq-1$, and an intersection of the line segment $\mu=1$ with $0\geq\nu\geq-1$ and the line segment between the optimizer $(\mu^{*},\nu^{*})$ and any point within the triangular region $\mathcal{B}_{2}$. With the joint concavity property of $T_{\mu,\nu}(Q_{1}||Q_{2};Q_{3})$ over $(\mu,\nu)\in\mathbb{R}^{2}$, we can conclude from these observations that
\begin{equation*}
\begin{aligned}
&\max_{(\mu,\nu)\in\mathbb{R}^{2}}T_{\mu,\nu}(Q_{1}||Q_{2};Q_{3})\\
&\;\;\;=\max_{(\mu,\nu)\in\mathcal{B}_{1}}T_{\mu,\nu}(Q_{1}||Q_{2};Q_{3})\\
&\;\;\;\geq\max_{1\geq\mu\geq0}-\log\sum_{a\in\mathcal{S}}Q_{1}^{\mu}(a)Q_{2}^{1-\mu}(a)=C(Q_{1}||Q_{2})\\
&\;\;\;\geq\max_{0\geq\nu\geq-1}-\log\sum_{a\in\mathcal{S}}Q_{1}^{1+\nu}(a)Q_{3}^{-\nu}(a)=C(Q_{1}||Q_{3})\\
&\;\;\;=\max_{(\mu,\nu)\in\mathcal{B}_{2}}T_{\mu,\nu}(Q_{1}||Q_{2};Q_{3}),
\end{aligned}
\end{equation*}
i.e., we obtain the equality in the case of the optimizer $(\mu^{*},\nu^{*})$ locating within the triangular region $\mathcal{B}_{1}$.
\end{IEEEproof}

\subsection{Proof of Proposition \ref{ld1}}
\begin{IEEEproof}
For all $\mu_{1}$, $\mu_{2}$, $\mu_{3}$, $\mu_{4}\in[0,1]$, we have upper bounds on the minimal error probability of the utility hypothesis test as
\begin{equation*}
\begin{aligned}
\alpha_{U}(&Y^{n},\phi_{s}^{n})\\
=&\,\sum_{y^{n}\in\mathcal{X}^{n}}\min_{u\in\{0,1\}}\left\{\sum_{p=0}^{1}p_{Y^{n}|u,p}(y^{n})p_{U,P}(u,p)\right\}\\
\leq&\,p_{\max}\sum_{y^{n}\in\mathcal{X}^{n}}\min_{u\in\{0,1\}}\left\{\sum_{p=0}^{1}p_{Y^{n}|u,p}(y^{n})\right\}\\
\overset{(a)}{\leq}&\,2p_{\max}\sum_{y^{n}\in\mathcal{X}^{n}}\frac{\prod_{u=0}^{1}\sum_{p=0}^{1}p_{Y^{n}|u,p}(y^{n})}{\sum_{u=0}^{1}\sum_{p=0}^{1}p_{Y^{n}|u,p}(y^{n})}\\
\leq&\,2p_{\max}\sum_{y^{n}\in\mathcal{X}^{n}}\sum_{\bar{p},\tilde{p}\in\{0,1\}}\frac{p_{Y^{n}|0,\tilde{p}}(y^{n})p_{Y^{n}|1,\bar{p}}(y^{n})}{p_{Y^{n}|0,\tilde{p}}(y^{n})+p_{Y^{n}|1,\bar{p}}(y^{n})}\\
\overset{(b)}{\leq}&\,2p_{\max}\sum_{y^{n}\in\mathcal{X}^{n}}p_{Y^{n}|0,0}^{\mu_{1}}(y^{n})p_{Y^{n}|1,0}^{1-\mu_{1}}(y^{n})\\
&\,\;\;\;\;\;\;\;\;\;\;\;\;\;\;\;\,+p_{Y^{n}|0,0}^{\mu_{2}}(y^{n})p_{Y^{n}|1,1}^{1-\mu_{2}}(y^{n})\\
&\,\;\;\;\;\;\;\;\;\;\;\;\;\;\;\;\,+p_{Y^{n}|0,1}^{\mu_{3}}(y^{n})p_{Y^{n}|1,0}^{1-\mu_{3}}(y^{n})\\
&\,\;\;\;\;\;\;\;\;\;\;\;\;\;\;\;\,+p_{Y^{n}|0,1}^{\mu_{4}}(y^{n})p_{Y^{n}|1,1}^{1-\mu_{4}}(y^{n}),
\end{aligned}
\end{equation*}
where $(a)$ and $(b)$ follow from the generalized (weighted) mean inequality \cite[Chapter 3]{bullen1987}. In particular, we have
\begin{equation*}
\begin{aligned}
\alpha_{U}(&Y^{n},\phi_{s}^{n})\\
\leq&\,2p_{\max}\min_{1\geq\mu_{1}\geq0}\sum_{y^{n}\in\mathcal{X}^{n}}p_{Y^{n}|0,0}^{\mu_{1}}(y^{n})p_{Y^{n}|1,0}^{1-\mu_{1}}(y^{n})\\
&\,+2p_{\max}\min_{1\geq\mu_{2}\geq0}\sum_{y^{n}\in\mathcal{X}^{n}}p_{Y^{n}|0,0}^{\mu_{2}}(y^{n})p_{Y^{n}|1,1}^{1-\mu_{2}}(y^{n})\\
&\,+2p_{\max}\min_{1\geq\mu_{3}\geq0}\sum_{y^{n}\in\mathcal{X}^{n}}p_{Y^{n}|0,1}^{\mu_{3}}(y^{n})p_{Y^{n}|1,0}^{1-\mu_{3}}(y^{n})\\
&\,+2p_{\max}\min_{1\geq\mu_{4}\geq0}\sum_{y^{n}\in\mathcal{X}^{n}}p_{Y^{n}|0,1}^{\mu_{4}}(y^{n})p_{Y^{n}|1,1}^{1-\mu_{4}}(y^{n})\\
\leq&\,8p_{\max}\max_{\bar{p},\tilde{p}\in\{0,1\}}\left\{\exp{\left(-C(p_{Y^{n}|1,\bar{p}}||p_{Y^{n}|0,\tilde{p}})\right)}\right\},
\end{aligned}
\end{equation*}
i.e., the error exponent of the hypothesis test on $U$ has the following lower bound:
\begin{equation*}
\begin{aligned}
\frac{1}{n}&\log\frac{1}{\alpha_{U}(Y^{n},\phi_{s}^{n})}\\
&\geq\frac{1}{n}\min_{\bar{p},\tilde{p}\in\{0,1\}}\left\{C(p_{Y^{n}|1,\bar{p}}||p_{Y^{n}|0,\tilde{p}})\right\}-\frac{\log8p_{\max}}{n}.
\end{aligned}
\end{equation*}

The proof of the lower bound on the error exponent of the privacy hypothesis test is similar and therefore is omitted.
\end{IEEEproof}

\bibliographystyle{IEEEtran}
\bibliography{References}

\end{document}